\documentclass{article}%
\usepackage{lineno}
\usepackage{xcolor,colortbl}
\usepackage{eurosym}
\usepackage[top=3cm, bottom=3cm, left=3cm, right=3cm]{geometry}
\usepackage{amsmath}
\usepackage{amsfonts}
\usepackage{amssymb}
\usepackage{graphicx}%
\usepackage{epsfig}
\usepackage{float}
\begin{document}

\title{\textbf{Static properties and Semileptonic transitions of lowest-lying double heavy baryons}}
\author{Zahra Ghalenovi  \footnote{$z_{-}ghalenovi@kub.ac.ir$}  and Masoumeh Moazzen Sorkhi\\Department of Physics, Kosar University of Bojnord, Iran
}

\maketitle

\begin{abstract}
The static properties and semileptonic decays of ground-state doubly heavy baryons are studied working in the framework of a non-relativistic quark model. Using a phenomenological potential model, we calculate the ground-state masses and magnetic moments of doubly heavy $ \Omega $ and $ \Xi $ baryons. In the heavy quark limit, we introduce a simple form of the universal Isgur-Wise function used as the transition form factor and then investigate the exclusive $ b \rightarrow c $  semileptonic decay widths and branching fractions for $ \frac{1}{2}\rightarrow \frac{1}{2} $ baryon transitions. Our obtained results are in agreement with other theoretical predictions. 
\end{abstract}
\section{Introduction}
The heavy baryons containing two heavy charm or bottom quarks is an interesting study that has been attracting during the last few years. In 2002, the lightest flavored charm-charm baryons was observed by the SELEX Collaboration in the analysis of decay modes
$ \Xi ^{+}_{cc} \rightarrow \Lambda _{c}^{+} K^{-}\pi ^{+} $ and $ \Xi ^{+}_{cc} \rightarrow \rho D^+ K^{-}$ \cite{Mattson2002, Ocherashvili2005}. They reported $M_{\Xi ^{++}_{cc}}=3518\pm1.7$ MeV. It was the first observation of double heavy baryons. Later, the LHCb collaboration reported the new mass values of the $ \Xi ^{++}_{cc}$ state as $ 3621.40\pm0.72\pm0.14\pm0.27$ MeV and $ 3620.6\pm1.5\pm0.4\pm0.3 $ MeV on different decay modes\cite{LHCb2017,LHCb2018}. Such important observations undoubtedly promote the investigation on the hadron spectroscopy and also on the weak decays of double heavy baryons. The beauty-charmed  and also double-beauty baryons are the different kinds of doubly heavy baryons. Considering the doubly heavy baryon states can be generated at the LHC or future high luminosity $ e^+e^- $ colliders, proposing theoretical studies on this subject are important. \\
A detailed understanding of semileptonic transitions necessitates an understanding of the interesting interplay between the weak and the strong interactions. The transition form factors and exclusive decay rates provide information about the heavy baryon structures and also provide another opportunity to measure the Cabibbo-Kobayashi-Maskawa (CKM) matrix elements. While there are different theoretical ~\cite{Li2021,Hsiao2020,Lu2021,Faustov2019,
Faustov2016,Migura2006,Azizi2012}  and experimental \cite{Ablikim2017,Ablikim2018,
Li2021a,Li2021b,Ammar2002} achievements on semileptonic decays of single heavy baryons but, for  doubly heavy baryons no experimental data are reported and only a limited number of theoretical calculations are available. \\
Some theoretical explanations are proposed to analysis the semileptonic weak decays of  double heavy baryons based on the QCD sum rules (QCDSR)\cite{Kiselev2001, Shi2020b}, covariant confined quark model \cite{Gutsche2019}, Bethe-Salpeter  equation \cite{Yu2019}, heavy quark spin symmetry (HQSS) \cite{Hernandez2008},  relativistic quark model (RQM) \cite{Lyubovitskij2003}, non-relativistic quark model (NRQM) \cite{Albertus2005,Albertus2008},   light-front quark model (LFQM)\cite{Wang2017} and heavy diquark effective theory (HDET)\cite{Qin2021,Shi2020a}. However, the studies on semileptonic decays presented by different approaches lead to essentially different values for the decay rates. \\

We aim to give a description of doubly heavy baryon properties within the non-relativistic hypercentral quark model. Firstly, we calculate the mass spectra and magnetic moments of the ground-state $ \Omega $ and $ \Xi $ baryons containing two heavy quarks (bottom or charm). Then, we study the semileptonic decays of doubly heavy baryons for specific decay modes and focus on the studies of exclusive  $ b\rightarrow c $ semileptonic decays of doubly heavy $ \Omega(J=1/2)$ and $ \Xi(J=1/2) $ baryons. We proceed our model close to the zero recoil point in which all weak transition form factors between double heavy baryons can be expressed through a single universal function $ \eta(\omega) $\cite{Faessler2009}. Introducing a phenomenological potential model we calculate the mass spectra and magnetic moments of double heavy baryons and then, using the obtained results we report our numerical results for the semileptonic decay widths of doubly heavy $ \Omega $ and $ \Xi $ baryons. \\
Many studies have presented various methods to solve the Schr\"{o}dinger equation of three-body baryonic system which are mostly long and complicated. The purpose of the present work is to simplify the solution of three-body problem using the hypercentral approach and calculate the semileptonic decay widths in the limit $ m_b, m_c\gg\Lambda_{QCD} $ to reduce the independent number of form factors and evaluate the decay widths in a simple way.  \\
 Our paper is structured as follows. We introduce our phenomenological potential model in section \ref{model} and simplify the three-body problem of the baryonic system using the hypercentral approach. In sections \ref{mass} and \ref{magnetic m}  we present our predictions for the ground-state masses and magnetic moments of doubly heavy baryons, respectively. In section \ref{semi decay} we simplify the transition form factors in the heavy quark limit and introduce a Isgur-Wise (IW) function to evaluate the semileptonic decay rates and branching ratios of doubly heavy baryons for $ b\rightarrow c $ transitions. We also calculate the averaged values of asymmetry parameters. Section \ref{Conclusions} contains our conclusions.

\section{Three body problem} \label{model}
In order to obtain the heavy baryon masses we need to calculate the energy eigenvalues and also the eigenfunctions of the baryon states. We study the doubly heavy baryons and therefore, the non-relativistic formalism is an appropriate method. We consider the baryon system as a bound state of three constituent quarks. Using the hypercentral constituent quark model \cite{Giannini:1998,Giannini:2} we can simplify the three-body problem of the baryon states. After removal of the center of mass coordinates, $ R $, the configurations of three quarks can be described by the Jacoobi coordinates, $ \rho $ and $ \lambda $,
\begin{equation}  \label{rhoo}%
\vec{\rho} = \frac{1}{\sqrt{2}}(\vec{r_1} - \vec{r_2}), \qquad
\vec{\lambda} = \frac{1}{\sqrt{6}}(\vec{r_1} + \vec{r_2} - 2
\vec{r_3})  
\end{equation}
such that
\begin{equation}
m_{\rho} = \frac{2 m_1 m_2}{m_1 + m_2}, \qquad m_{\lambda} =
\frac{3 m_3 (m_1 + m_2)}{2 (m_1 + m_2 + m_3)}  
\end{equation}
Here $m_1$, $m_2$ and $m_3$ are the constituent quark masses.
Instead of $\rho$ and $\lambda$ , one can introduce the
hyperspherical coordinates, which are given by the angles
$\Omega_{\rho}=(\theta_{\rho},\phi_{\rho})$ and $\Omega_{\lambda}=(\theta_{\lambda},\phi_{\lambda})$ together with the
hyperradius $x$, and the hyperangle, $\zeta$ defined in terms of the absolute values of   $\rho$ and $\lambda$  
by 
\begin{equation}
x = \sqrt{\rho^2 + \lambda^2}, \qquad \xi = \arctan
(\sqrt{\frac{\rho}{\lambda}}).
\end{equation}

Therefore, the Hamiltonian of three-body system is given by
\begin{equation}
H=\frac{p^{2}_{\rho}}{2m_{\rho}}+\frac{p^{2}_{\lambda}}{2m_{\lambda}}+V(x).
\end{equation}

and the kinetic energy operator takes the form ($ \hbar=c=1 $)
\begin{equation}    \label{laplace}
-\left(\dfrac{\Delta_{\rho}}{2m_{\rho}}+\dfrac{\Delta_{\lambda}}{2m_{\lambda}}\right)=-\dfrac{1}{2\mu}(\frac{d^{2}}{dx^{2}}+\frac{5}{x}\frac{d}{dx}-\frac{L^{2}(\Omega_{\rho},\Omega_{\lambda},\xi)}{x^{2}}),
\end{equation}

where, $ \mu $ is the reduced mass 

\begin{equation}
\mu=\dfrac{2m_{\rho}m_{\lambda}}{m_{\rho}+m_{\lambda}},
\end{equation}

and $\Omega_{\rho}  $ and $ \Omega_{\lambda}  $ are the angels of hyperspherical coordinates. In the hypercentral approach, the three-quark interaction is assumed to be hypercentral, that is the energy of the system depends only on the quarks distance $x$.

The eigenfuctions of $L^{2}$ are hyperspherical harmonics
\begin{equation}
L^{2}(\Omega_{\rho},\Omega_{\lambda},\xi)Y_{[\gamma],l_{\rho},l_{\lambda}}(\Omega_{\rho},\Omega_{\lambda},\xi)=\gamma(\gamma+1)Y_{[\gamma],l_{\rho},l_{\lambda}}(\Omega_{\rho},\Omega_{\lambda},\xi),
\end{equation}

where $\gamma$ is the grand angular quantum number given by $\gamma=2n+l_\rho+l_\lambda$  ; $l_\rho$ and $l_\lambda$  are the angular momenta associated with the  $\rho$ and $\lambda$ variables and $n$ is a non-negative integer number. The wave function of any system containing three particles can be expanded in the hyperspherical harmonic basis:

\begin{equation}   \label{Psi}
\Psi(\rho,\lambda)=\Sigma_{\gamma,l_{\rho},l_{\lambda}}  N_{\gamma}\psi_{\nu,\gamma}(x)Y_{[\gamma],l_{\rho},l_{\lambda}}(\Omega_{\rho},\Omega_{\lambda},\xi).  
\end{equation}

where $\nu$ determines the number of the nodes of the wave function. The hyperradial wave funtion $ \psi_{\nu,\gamma}(x) $ is obtained as a solution of the hyperradial equation \cite{Giannini:2,Narodetskii,Ghalenovi:2011,Ghalenovi:2014}

\begin{equation}\label{Schro 2}
[\frac{d^2}{dx^2}+\frac{5}{x}\frac{d}{dx}-\frac{\gamma(\gamma+4)}{x^2}]\psi_{\nu,\gamma}(x)=-2\mu[E_{\nu,\gamma}
-V(x)]\psi_{\nu,\gamma}(x).
\end{equation}
 For the ground-state doubly heavy baryons we have $ \gamma=\nu=0 $. Here, $\mu$ is a free parameter with dimension of mass \cite{Narodetskii,Ghalenovi:2011,Ghalenovi:2014}. We consider the Coulomb-plus-linear potential known as the Cornell potential and given by

 \begin{equation} \label{v(x)}
 V(x)=-\frac{\tau}{x}+\beta x.
 \end{equation}
 
Here, $x$ is the hyper radius  and $\tau$ and $\beta$ are constant. In the present work, we solve the Schr\"{o}dinger equation \ref{Schro 2} numerically. \\
In our model, the spin dependent interaction is treated as perturbation. We make the following model for the spin-spin hyperfine interaction

\begin{equation}   \label{Hs}
V_{spin}=\frac{A_{s}}{m_\rho m_\lambda}\frac{e^{\frac{-x}{x_{0}}}}{x x_{0}^2}\Sigma_{i<j}(\overrightarrow{s_{i}}.\overrightarrow{s_j}),
\end{equation}
where $\overrightarrow{s_i}$ is the spin operator of the $i^{th}$  quark and $ A_s $ and $ x_0 $ are constants. We have neglected the isospin dependent potential. The isospin mass splittings for doubly heavy baryons have been studied in Ref. \cite{Brodsky2011} with the results 

\begin{eqnarray}
m_{\Xi^{++}_{cc}}-m_{\Xi^{+}_{cc}}=(1.5\pm 2.7)MeV, \nonumber \\
m_{\Xi^{+}_{bc}}-m_{\Xi^{0}_{bc}}=(-1.5\pm0.9)MeV,\\
m_{\Xi^{-}_{bb}}-m_{\Xi^{0}_{bb}}=(6.3\pm1.7)MeV. \nonumber
\end{eqnarray}

Thus, the isospin splittings are at most a few MeV and can be neglected in our calculations.  

\section{Baryon spectrum} \label{mass}

From the experimental side only the mass  of the $ \Xi_{cc} $ state has been measured. Other than the double-charm $ \Xi_{cc}$  baryons, the masses of other 
doubly heavy baryons are not known yet and may be measured in the future. 
Recently, several experimental efforts have been made on the exclusive channels
$ \Xi^0_{bc}\rightarrow D^0 p K^-$~\cite{lhcb1} and $ \Xi^0_{bc}\rightarrow
\Xi_c^+ \pi^-$~\cite{lhcb2}  to search for the beauty-charm baryons, but no signals were
observed. Recently, an inclusive decay channel
$ \Xi_{bc}\rightarrow \Xi_{cc}^{++}+X$ has been proposed~\cite{qqin} to search for the $ \Xi_{bc}$ baryons where, $ X $ stands for all the possible particles. The theoretical results on the mass spectrum of doubly heavy baryons would be also helpful and can be tested in the near future.\\
In theory, the mass of the baryon is simply given by the energy of the baryonic system.
One can calculate the baryon masses by the sum of the quark masses, energy eigenvalues and perturbative hyperfine interaction $\left\langle V_{spin}\right\rangle $  as 

\begin{equation}\label{Mass}
M_{baryon}=m_{1}+m_{2}+m_{3}+E+\left\langle V_{spin} \right\rangle,
\end{equation}

 The all model parameters (listed in table \ref{tab:parameters}) are taken from our previous work \cite{Ghalenovi:2014} in which we have studied the properties of charm and bottom heavy baryons in a quark model. Our results for the ground-state baryons are listed in table \ref{tab:mass1} and compared to those reported within a relativistic quark model \cite{Ebert2004}, non-relativistic quark model \cite{Albertus2010,Roberts2008,Yoshida2015} and QCD sum rules \cite{Azizi2013}. Our evaluation for $ M_{\Xi_{cc}} $ is 58  MeV higher than the experimental value reported by LHCb collaboration \cite{LHCb2017}. Our predicted masses are in good agreement with the ones reported by Refs. \cite{Roberts2008}. For the $ \Omega_{bb}^* $ our mass prediction is the same as one reported by Ref.\cite{Roberts2008}.

 \begin{table} [h] 
 \caption{Quark-model parameters.  $ q $ refers to the light quarks $ u $ and $ d $.}
 \label{tab:parameters}
 \begin{center}
\begin{tabular}{llll} \hline \hline  
Parameter&Value&Parameter&Value\\ \hline           
$m_{q}$  & 330 $MeV$&$\tau$  & 4.59\\ 
$m_{s}$  & 469  $MeV$&$\beta$ & 1.61 $fm^{-2}$\\
$m_{c}$  & 1600 $MeV$&$A_s$&67.4 \\
$m_{b}$  & 4908 $MeV$ &$x_0$& 2.87 $fm$\\
$ \mu $  &0.884 $fm^{-1}$&\\

 \hline

\end{tabular}
  \end{center}
   \end{table}

\begin{table} [h]
\caption{Masses of the ground states of double heavy baryons (in GeV). $ ^* $ refers to the $ s=\frac{3}{2} $ baryons.}
\label{tab:mass1}
 \begin{center}
{\begin{tabular}{llllllll} \hline \hline  
Baryon& Content&Our results& \cite{Ebert2004}&\cite{Albertus2010} &\cite{Roberts2008}& \cite{Yoshida2015}&\cite{Azizi2013}\\  \hline
$ \Xi_{cc} $&$ qcc $&  3.679& 3.620&3.613& 3.676&3.685&\\
$ \Xi_{cc}^* $&$ qcc $&  3.763&3.727&3.707&3.753&3.754&3.690 \\
$ \Xi_{bc} $&$ qbc $&  7.003&6.933&6.928& 7.020& &\\
$ \Xi_{bc}^* $  &$ qbc $&  7.056&6.980&6.996&7.078& & 7.250\\
$ \Xi_{bb} $  &$ qbb $& 10.325&10.202&10.198& 10.340 &10.314&\\
$ \Xi_{bb}^* $  &$ qbb $& 10.350& 10.237&10.237&10.367&10.339&10.400 \\
$ \Omega_{cc} $&$ scc $& 3.830 & 3.778&3.712&3.815&3.832& \\
$ \Omega_{cc}^* $&$ scc $& 3.891 & 3.872&3.795&3.876&3.883&3.780\\
$ \Omega_{bc} $&$ sbc $& 7.149  & 7.088&7.013& 7.147&&\\
$ \Omega_{bc}^* $  &$ sbc $& 7.187 & 7.130&7.075&7.191&&7.300\\
$ \Omega_{bb} $  &$ sbb $& 10.466 & 10.359&10.269&10.456&10.447&\\
$ \Omega_{bb}^* $  &$ sbb $& 10.486  & 10.389&10.307&10.486&10.467&10.500\\
\hline

\end{tabular}}
 \end{center}
\end{table}

For the mass splittings between $ J=\frac{3}{2} $ and corresponding $ J=\frac{1}{2} $ doubly heavy baryons, our calculations show $ M_{\Xi_{cc}^*}-M_{\Xi_{cc}}=84$ MeV , $ M_{\Xi_{bc}^*}-M_{\Xi_{bc}}=53$ MeV, $ M_{\Xi_{bb}^*}-M_{\Xi_{bb}}=25$ MeV, $ M_{\Omega_{cc}^*}-M_{\Omega_{cc}}=61$ MeV, $ M_{\Omega_{bc}^*}-M_{\Omega_{bc}}=38$ MeV and $ M_{\Omega_{bb}^*}-M_{\Omega_{bb}}=20$ MeV. The obtained results of Ref. \cite{Roberts2008} show  $ M_{\Xi_{cc}^*}-M_{\Xi_{cc}}77=$ MeV , $ M_{\Xi_{bc}^*}-M_{\Xi_{bc}}=58$ MeV, $ M_{\Xi_{bb}^*}-M_{\Xi_{bb}}=27$ MeV, $ M_{\Omega_{cc}^*}-M_{\Omega_{cc}}=61$ MeV, $ M_{\Omega_{bc}^*}-M_{\Omega_{bc}}=44$ MeV and $ M_{\Omega_{bb}^*}-M_{\Omega_{bb}}=30$ MeV which are very close to our calculations, respectively. For doubly bottom baryons our evaluated mass splittings $ M_{\Xi_{bb}^*}-M_{\Xi_{bb}} $ and $ M_{\Omega_{bb}^*}-M_{\Omega_{bb}}$ are the same as those obtained by Ref. \cite{Yoshida2015}.\\
For the mass splittings between the double heavy baryons containing two identical heavy quarks and corresponding bottom-charm baryons we find
\begin{eqnarray}
m_{\Xi_{bc}}-m_{\Xi_{cc}}=3324 MeV, \nonumber \\
m_{\Xi_{bb}}-m_{\Xi_{bc}}=3322 MeV,\\ \nonumber 
m_{\Omega_{bc}}-m_{\Omega_{cc}}=3319 MeV,\\
m_{\Omega_{bb}}-m_{\Omega_{bc}}=3317 MeV. \nonumber
\end{eqnarray}
which all the splittings are close to the mass difference of the bottom and charm quarks
\begin{equation}
m_b-m_c=3309 MeV.
\end{equation}
In the case of $ m_{\Xi_{bb}}-m_{\Xi_{bc}} $, our result are the same as mass splittings reported in Refs. \cite{Albertus2010} and \cite{Roberts2008}.

It is also interesting to pointed out the mass difference $ \Delta_M=M_{\Omega_{QQ}}-M_{\Xi_{QQ}}$ between $ \Omega $ and corresponding $ \Xi $ doubly heavy baryons. In our calculations it takes the value of $ \Delta_M=142 \sim151 MeV$ which is close to the $ \Delta_M=155\sim158 MeV$ reported by Ref. \cite{Ebert2004}.

 \section{Magnetic moments} \label{magnetic m}
 The orbital part of the magnetic moment is defined in terms of the velocities $ \overrightarrow{v} $ of the quarks. In the present work we study the ground-state heavy baryons and then, the orbital magnetic moments vanishes and the magnetic moment of the baryon is entirely given by the spin contribution. Then, the magnetic moment operator is given as

\begin{equation} \label{magnetic moB}
\overrightarrow{\mu}=\sum_i \mu_i \overrightarrow{\sigma}_i,
\end{equation}

where
\begin{equation} \label{magnetic mo}
\mu_i=\dfrac{e_{i}}{2m_{i}},
\end{equation}
 
and $ e_i $ and $ \sigma_i $ represent the charge and spin of the $i^{th}$ quark respectively ($ s_i=\dfrac{\sigma_i}{2}) $. 
The magnetic moment of the baryon is obtained in terms of its constituent quark magnetic moments ($ \mu_i $). By sandwiching the magnetic moment operators between the appropriate baryon wave functions we get
 \begin{equation} \label{magnetic mo}
\begin{aligned}
\mu=& \left\langle B_{sf}\mid \mu_i \overrightarrow{\sigma}_i \mid B_{sf} \right\rangle
\\
=& \left\langle B_{sf}\mid \dfrac{e_{Q_1}}{m_{Q_1}}(\overrightarrow{s}_{Q_1})_z+\dfrac{e_{Q_2}}{m_{Q_2}}(\overrightarrow{s}_{Q_2})_z+\dfrac{e_{q}}{m_{q}}(\overrightarrow{s}_{q})_z \mid B_{sf} \right\rangle,
\end{aligned}
\end{equation}

 where $Q $ and $ q $ refer to the heavy and light quarks, respectively and $ \mid B_{sf}> $ is the spin-flavor wave function of the baryon. The matrix elements \ref{magnetic mo} are evaluated as follows
 
 \begin{eqnarray}
 \mu_{B(J=1/2)}\longrightarrow \dfrac{2}{3}\left(\dfrac{e_{Q_1}}{2m_{Q_1}}+\dfrac{e_{Q_2}}{2m_{Q_2}}-\dfrac{1}{2}\dfrac{e_{q}}{2m_{q}}\right), \nonumber  \\
 \mu_{B(J=3/2)}\longrightarrow \dfrac{e_{Q_1}}{2m_{Q_1}}+\dfrac{e_{Q_2}}{2m_{Q_2}}+\dfrac{e_{q}}{2m_{q}}.  \qquad  \quad
 \end{eqnarray}
 
 With $ m_b\gg m_u,m_d,m_s $, the contribution from the $ b $ quark is negligible compared to the ones of the light quarks. This is also true to a lesser extent for the $ c $ quark.
 
\begin{table} [H] 
\begin{center}
\caption{Spin-flavor wave functions of  $ \Xi $  doubly heavy baryons and their magnetic moment in terms of the constituent quarks magnetic moments. $ ^* $ refers to the $ s=\frac{3}{2} $ baryons.}  \label{tab:spin-flavor}
\begin{tabular}[c]{c|c|c} \hline 
Baryon&Spin-flavour wave function & $ \mu_B $ \\  \hline 
$\Xi_{cc}^{++}$&$\frac{2}{\sqrt{6}}(2u_-c_+c_+-c_-u_+c_+-u_+c_-c_++2c_+u_-c_+-c_+c_-u_+$&\\
&$-c_-c_+u+-c_+u_+c_--u_+c_+c_-+2c_+c_+u_-)   $ & $\frac{4}{3}\mu_c-\frac{1}{3}\mu_u $\\
$\Xi_{cc}^{+} $          & $ \frac{2}{\sqrt{6}}(2d_-c_+c_+-c_-d_+c_+-d_+c_-c_++2c_+d_-c_+-c_+c_-d_+$&\\
&$-c_-c_+d+-c_+d_+c_--d_+c_+c_-+2c_+c_+d_-)   $ & $\frac{4}{3}\mu_c-\frac{1}{3}\mu_d $\\  
$\Xi_{bb}^{0} $          & $ \frac{2}{\sqrt{6}}(2u_-b_+b_+-b_-u_+b_+-u_+b_-b_++2b_+u_-b_+-b_+b_-u_+$&\\
&$-b_-b_+u+-b_+u_+b_--u_+b_+b_-+2b_+b_+u_-)   $ & $\frac{4}{3}\mu_b-\frac{1}{3}\mu_u $\\     
$\Xi_{bb}^{-} $          & $ \frac{2}{\sqrt{6}}(2d_-b_+b_+-b_-d_+b_+-d_+b_-b_++2b_+d_-b_+-b_+b_-d_+$&\\
&$-b_-b_+d+-b_+d_+b_--d_+b_+b_-+2b_+b_+d_-)   $ & $\frac{4}{3}\mu_b-\frac{1}{3}\mu_d $\\  
$\Xi_{bc}^{+} $          & $ \frac{-1}{6}(b_+c_-u_++c_+b_-u_++u_+c_-b_++u_+b_-c_+-2b_+u_-c_+$\\
&  $-2c_+u_-b_++b_-c_+u_++c_-b_+u_+-2u_-c_+b_+-2u_-b_+c_+ $ & $\frac{2}{3}\mu_b+\frac{2}{3}\mu_c-\frac{1}{3}\mu_u $\\ 
& $+b_-u_+c_++c_-u_+b_+-2b_+c_+u_--2c_+b_+u_-+u_+c_+b_- $ &\\ 
&  $ +u_+b_+c_-+b_+u_+c_-+c_+u_+b_-)$ &\\  
$\Xi_{bc}^{0} $          & $ \frac{-1}{6}(b_+c_-d_++c_+b_-d_++d_+c_-b_++d_+b_-c_+-2b_+d_-c_+$\\
&  $-2c_+d_-b_++b_-c_+d_++c_-b_+d_+-2d_-c_+b_+-2d_-b_+c_+ $ & $\frac{2}{3}\mu_b+\frac{2}{3}\mu_c-\frac{1}{3}\mu_d $\\ 
& $+b_-d_+c_++c_-d_+b_+-2b_+c_+d_--2c_+b_+d_-+d_+c_+b_- $ &\\ 
&  $ +d_+b_+c_-+b_+d_+c_-+c_+d_+b_-)$ &\\ 
$\Xi_{cc}^{*++} $  & $\frac{1}{\sqrt{3}}(c_+c_+u_++c_+u_+c_++u_+c_+c_+)$ &$2\mu_c+\mu_u $\\
$\Xi_{cc}^{*+} $  & $\frac{1}{\sqrt{3}}(c_+c_+d_++c_+d_+c_++d_+c_+c_+)$ &$2\mu_c+\mu_d $\\
$\Xi_{bb}^{*0} $  & $\frac{1}{\sqrt{3}}(b_+b_+u_++b_+u_+b_++u_+b_+b_+)$ &$2\mu_b+\mu_u $\\
$\Xi_{bb}^{*-} $  & $\frac{1}{\sqrt{3}}(b_+b_+d_++b_+d_+b_++d_+b_+b_+)$ &$2\mu_b+\mu_d $\\ 
$\Xi_{bc}^{*+} $  & $\frac{1}{\sqrt{6}}(b_+c_+u_++c_+b_+u_++b_+u_+c_+$\\
& $+u_+b_+c_++c_+u_+b_++u_+c_+b_+)$ &$\mu_b+\mu_c+\mu_u $\\
$\Xi_{bc}^{*0} $  & $\frac{1}{\sqrt{6}}(b_+c_+d_++c_+b_+d_++b_+d_+c_+$\\
& $+d_+b_+c_++c_+d_+b_++d_+c_+b_+)$ &$\mu_b+\mu_c+\mu_d $\\
 \hline
\end{tabular}
\end{center}
\end{table}

The spin-flavor wave functions of doubly heavy $ \Xi $ baryons and their magnetic moments in terms of the constituent quarks magnetic moments are presented in table \ref{tab:spin-flavor}. In the same way, if the light ($ u$ and $d$) quarks   be replaced by the $ s $  quark, we can get the results for the corresponding $ \Omega $  doubly heavy baryons.  In table \ref{magnetic} we present our numerical results for the magnetic moments of doubly heavy baryons in terms of the nuclear magneton $ \mu_{N}$  and compare them to the ones obtained by different approaches \cite{Albertus2008,Patel2008,Dhir2013,Dhir2009}. 
 
\begin{table}
\caption{Magnetic moments of the  $ \Xi $ and $\Omega $  doubly heavy baryons (in $ \mu_{N} $) . \label{magnetic}}
 \begin{center}
{\begin{tabular}{cccccc} \hline \hline
Baryon& Our results&\cite{Albertus2008}&\cite{Patel2008}&\cite{Dhir2013}&\cite{Dhir2009}  \\ \hline
$ \Xi_{cc}^{++} $& -0.110&-0.208&-0.137&&\\
$ \Xi_{cc}^{+} $&0.836 &0.785&0.859&&\\
$ \Xi_{cc}^{*++}$&2.676 &2.670&2.749&&2.590\\
$ \Xi_{cc}^{*+}$& -0.165&-0.311 &-0.168&&-0.200\\
$ \Xi_{bc}^+ $&-0.413 &-0.475&-0.400&-0.387&\\
$ \Xi_{bc}^0 $&0.533 & 0.518&0.476&0.499&\\
$ \Xi_{bc}^{*+} $ &2.222 &2.270  &2.052&&2.011\\
$ \Xi_{bc}^{*0} $  &-0.620  & -0.712&-0.567& &-0.551\\
$ \Xi_{bb}^0 $&-0.716 &-0.742&-0.656&-0.665&\\
$ \Xi_{bb}^- $&0.230 &0.251 &0.190&0.208&\\
$ \Xi_{bb}^{*0} $ &1.767 &1.870 &1.576& &1.596\\
$ \Xi_{bb}^{*-} $  &-1.074 &-1.110 &-0.951&&-0.984\\
$ \Omega_{cc}^+ $&0.743 &0.635 &0.783&&\\
$ \Omega_{cc}^{*+} $& 0.114&0.139 &0.121&&0.120\\
$ \Omega_{bc}^0 $&0.440 &0.368 &0.396&0.399&\\
$ \Omega_{bc}^{*0} $  &-0.339 &-0.261 &-0.316&&-0.279\\
$ \Omega_{bb}^{-} $& 0.137&0.101 &0.109&0.111& \\
$ \Omega_{bb}^{*-} $&-0.794 &-0.662&-0.711& &-0.703\\
 \hline
\end{tabular}}
\end{center}
\end{table}
 
\section{Semileptonic $ B\rightarrow B^{\prime}l\bar{\nu} $ decay widhts and branching ratios  } \label{semi decay}

In semileptonic decays a hadron decays weakly into another hadron with the emission of
a lepton and a neutrino. In this section we would like to study different doubly heavy $ B(1/2^+)\rightarrow B^{\prime}(1/2^+) $ baryon semileptonic decays for $ b\rightarrow c $ transitions where, doubly heavy baryon $ B $ with $ J=1/2 $ decays into another doubly heavy baryon $ B^{\prime} $ with the same spin. The transition matrix element for the semileptonic decay is

\begin{equation}
T=\dfrac{G_F}{\sqrt{2}}V_{cb}J_{\mu}\mathcal{L}^{\mu}
\end{equation}

with $G_F $ and $ V_{bc} $  representing the Fermi Coupling constant and CKM matrix element, respectively. $ J_{\mu} $ is the flavor changing (weak) hadronic current and      $ \mathcal{L}^{\mu}$ is the leptonic current given as 

\begin{equation}
J_{\mu} = V_{\mu}-A_{\mu}= \bar{\psi}^c\gamma_{\mu}(I-\gamma_5)\psi^b,     \qquad\mathcal{L}^{\mu}=\bar{l}\gamma^{\mu}(1-\gamma_5)\nu_l
\end{equation}

where $ \psi^{b(c)} $ refers to the  bottom (charm) quark field. $ V_{\mu}\equiv\bar{\psi}^c\gamma_{\mu}\psi^b $ and $ A_{\mu}\equiv\bar{\psi}^c\gamma_{\mu}\gamma_5\psi^b $ are the vector and axial-vector weak currents, respectively. The leptonic current of the semileptonic decay can be calculated from first principles of Quantum mechanics but hadronic matrix element, $ H_{\mu} $, is not. The hadron matrix elements can be parametrized in terms of six form factors as
\begin{equation}
\begin{aligned}
H_{\mu}&= 
\left\langle B^{\prime}(p^{\prime},s^{\prime})\mid V_{\mu}-A_{\mu}\mid B(p,s)\right\rangle=\left\langle B^{\prime}(p^{\prime},s^{\prime})\mid \bar{\psi}^c\gamma_{\mu}(I-\gamma_5)\psi^b\mid B(p,s)\right\rangle
\\
=&\bar{u}^ {\prime}(p^{\prime},s^{\prime})\left\lbrace \gamma_{\mu}(F_1(\omega)-\gamma_5G_1(\omega))+v_{\mu}(F_2(\omega)-\gamma_5G_2(\omega)) \right. \\
& \left. +v^{\prime}_{\mu}(F_3(\omega)-\gamma_5G_3(\omega))\right\rbrace u(p,s)
\end{aligned}
\end{equation}

where $ \mid B(p,s)> $ and $ \mid B^{\prime}(p^{\prime},s^{\prime})> $ represent the initial and final baryons. $ u(p,s) $ and $ u^{\prime}(p^{\prime},s^{\prime}) $ are dimensionless Dirac spinors, normalized as $ \bar{u}u=1 $, and $ v_{\mu}=p_{\mu}/m_B (v^{\prime}_{\mu}=p^{\prime}_{\mu}/m_B^{\prime})$ is the four velocity of the initial $ B $ (final $ B^{\prime} $) baryon.

 The differential decay rates from transversely $ \Gamma_T $ and longitudinally $ \Gamma_L $ polarized W's, are given as

\begin{equation}
\dfrac{d\Gamma}{d\omega}=\dfrac{d\Gamma_L}{d\omega}+\dfrac{d\Gamma_T}{d\omega}
\end{equation}

Neglecting the lepton masses, the differential decay rates $ \Gamma_T $ and $ \Gamma_L $ for the case $ B^{1/2}\rightarrow B^{\prime 1/2} $ are given as  \cite{Albertus2005,Albertus2008} 

\begin{equation}\label{transverss}
\dfrac{d\Gamma_T}{d\omega}=\frac{G_F^2\vert V_{cb}\vert ^2M_{B^{\prime}}^3}{12\pi^3}\sqrt{\omega^2-1}q^2\lbrace(\omega-1)\vert F_1(\omega)\vert^2+(\omega+1)\vert G_1(\omega)\vert^2\rbrace,
\end{equation}

and

\begin{equation} \label{longitudd}
\dfrac{d\Gamma_L}{d\omega}=\frac{G_F^2\vert V_{cb}\vert ^2M_{B^{\prime}}^3}{24\pi^3}\sqrt{\omega^2-1}\lbrace(\omega-1)\vert \mathcal{F}^V(\omega)\vert^2+(\omega+1)\vert \mathcal{F}^A(\omega)\vert^2\rbrace,
\end{equation}

where

\begin{equation}
\begin{aligned}
\mathcal{F}^{V,A}(\omega)=&\left[ (m_B\pm m_{B^{\prime}})F_1^{V,A}(\omega)+(1\pm\omega) \left( m_{B^{\prime}}F_2^{V,A}(\omega)+m_BF_3^{V,A}(\omega)\right) \right] ,\\&
F_j^V\equiv F_j(\omega), \quad F^A_j\equiv G_j(\omega), \quad j=1,2,3
\end{aligned}
\end{equation}

$\omega=v.v^{\prime} $ is the velocity transfer and $ q^2=(p-p^{\prime})^2=m_B^2+m^2_{B^{\prime}}-2m_B m_{B^{\prime}}\omega$ is the momentum transfer squared between the initial and final baryon. 
The form factors are functions of $ \omega $ and $ q^2 $. In the decay $ \omega $ ranges from $ \omega=1 $ corresponding to zero recoil of the final baryon to a maximum $ \omega_{max}=(m_B^2+m^2_{B^{\prime}})/(2m_B m_{B^{\prime}})$. \\
Integrating over the parameter $ \omega $, we can obtain the total decay width 

\begin{equation} \label{total gamma}
  \Gamma=\int^{\omega_{max}}_1 d\omega\dfrac{d\Gamma}{d\omega}=\int^{\omega_{max}}_1 d\omega(\dfrac{d\Gamma_L}{d\omega}+\dfrac{d\Gamma_T}{d\omega}).
 \end{equation} 
We can also calculate the polar angle distribution  \cite{Albertus2005,Albertus2008}

\begin{equation} \label{longitud}
\dfrac{d^2\Gamma_L}{d\omega \, dcos\theta}=\dfrac{3}{8}\left(\dfrac{d\Gamma_T}{d\omega}+2\dfrac{d\Gamma_L}{d\omega}\right) \left\lbrace 1+2\alpha^{\prime} cos\theta+\alpha^{\prime\prime} cos^2\theta\right\rbrace 
\end{equation}

where $ \theta $ is the angle between the four momenta of the final baryon and the final lepton measured in the off-shell $ W $ rest frame.  $ \alpha $ and $ \alpha^{\prime \prime} $ are asymmetry parameters given by 
\begin{equation}
\alpha^{\prime}=-\frac{G_F^2\vert V_{cb}\vert ^2M_{B^{\prime}}^3}{6\pi^3}\dfrac{q^2(\omega^2-1)F_1(\omega)G_1(\omega)}{d\Gamma_T/d\omega+2d\Gamma_L/d\omega},
\end{equation}

and

\begin{equation}
\alpha^{\prime\prime}=\dfrac{d\Gamma_T/d\omega-2d\Gamma_L/d\omega}{d\Gamma_T/d\omega+2d\Gamma_L/d\omega}
\end{equation}

on averaging over $ \omega $ we get

\begin{equation}
<\alpha^{\prime}>=-\frac{G_F^2\vert V_{cb}\vert ^2}{6\pi^3} \dfrac{M^3_{B^{\prime}}}{\Gamma_T} \dfrac{\int ^{\omega_{max}}_1 q^2(\omega^2-1)F_1(\omega)G_1(\omega)d\omega}{1+2R_{L/T}}, 
\end{equation}

\begin{equation}
 <\alpha^{\prime \prime}>=\dfrac{1-2R_{L/T}}{1+2R_{L/T}}, \quad  \quad R_{L/T}=\dfrac{\Gamma_L}{\Gamma_T}.
\end{equation}

To study the semileptonic transitions of baryons we need the form factors which can be parametrized in different ways. Some earlier works  \cite{Faessler2009,Georgi1990, Carone1991,Flynn2008} have simplified the transition form factors using the different methods. The authors of Ref. \cite{Faessler2009} have studied the form factors and semileptonic decays of doubly heavy baryons using a manifestly Lorentz covariant constituent three-quark model. In the heavy quark limit and close to zero recoil point, the expressions for the rates can be  simplified considerably and the weak transition form factors between doubly heavy baryons can be expressed by a single  IW function $\eta(\omega)$ ~\cite{Faessler2009,Isgur1991,Ebert20066} 
 
\begin{eqnarray} \label{formfactors}
F_1(\omega)=G_1(\omega)=\eta(\omega), \quad \quad \quad \quad \quad\quad\\
F_2(\omega)=F_3(\omega)=G_2(\omega)=G_3(\omega)=0 \nonumber.
\end{eqnarray}

 The universal function $\eta(\omega)$ depends on the kinematical parameter $ \omega $  \cite{Faessler2009} 
\begin{equation} \label{eta}
\eta(\omega)=exp\left( -3(\omega-1)\frac{m_{cc}^2}{\Lambda_B^2}\right) 
\end{equation}  

with $m_{cc}=2m_c$ for the $ bc\rightarrow cc $  weak transitions. The cut-off parameter $ \Lambda_B $ is an adjustable parameter related to the size of a baryon and its value has
been determined as $ 2.5\leqslant \Lambda_B\leqslant 3.5$ GeV \cite{Faessler2009}. The values of the size parameter $ \Lambda_B $  are fixed using data on fundamental properties of mesons and baryons such as leptonic decay constants, magnetic moments and radii. By replacement $ m_{cc}\rightarrow m_{bb} $ in the IW function one can obtain the results for the $ bb\rightarrow bc $ transitions. Close to zero recoil, the IW functions for $ bb\rightarrow bc $ and $ bc\rightarrow cc $ transitions explicitly contain the flavor factors $ m_{cc}$ and $ m_{bb} $, respectively and there exists only spin symmetry. There is no dependence on the light quark masses. At zero recoil ($ \omega=1 $) there exists a spin-flavor symmetry and $ \eta(1)=1 $ means that $ bb\rightarrow bc $ and $ bc\rightarrow cc $ transitions are identical. \\

Using equation \ref{formfactors} we get the following relations for the differential decay rates:
 
\begin{equation}\label{transvers2}
\dfrac{d\Gamma_T}{d\omega}=
\frac{G_F^2\vert V_{cb}\vert ^2M_{B^{\prime}}^3}{6\pi^3}q^2\omega\sqrt{\omega^2-1} \eta^2(\omega),
\end{equation}

and

\begin{equation} \label{longitud2}
\begin{aligned}
&\dfrac{d\Gamma_L}{d\omega}=  \frac{G_F^2\vert V_{cb}\vert ^2M_{B^{\prime}}^3}{24\pi^3}\sqrt{\omega^2-1} \\&
\times\left[ (\omega-1)(m_B+m_{B^{\prime}})^2+(\omega+1)(m_B-m_{B^{\prime}})^2 \right]  \eta^2(\omega).
\end{aligned}
\end{equation}
 
Using the obtained masses listed in table \ref{tab:mass1} and Eqs. \ref{total gamma},  \ref{transvers2} and \ref{longitud2} one can calculate the semileptonic decay rates of doubly heavy  $ \Xi $ and $ \Omega $ baryons. For that purpose we need to fix a value for $ \vert V_{cb}\vert $. We take $\vert V_{cb}\vert=0.0422$ \cite{PDG2018}. We neglect the mass difference between the $ u $ and $ d $ light quarks.  \\

\begin{figure}[htbp]
\centering
\includegraphics[width=0.4\textwidth]{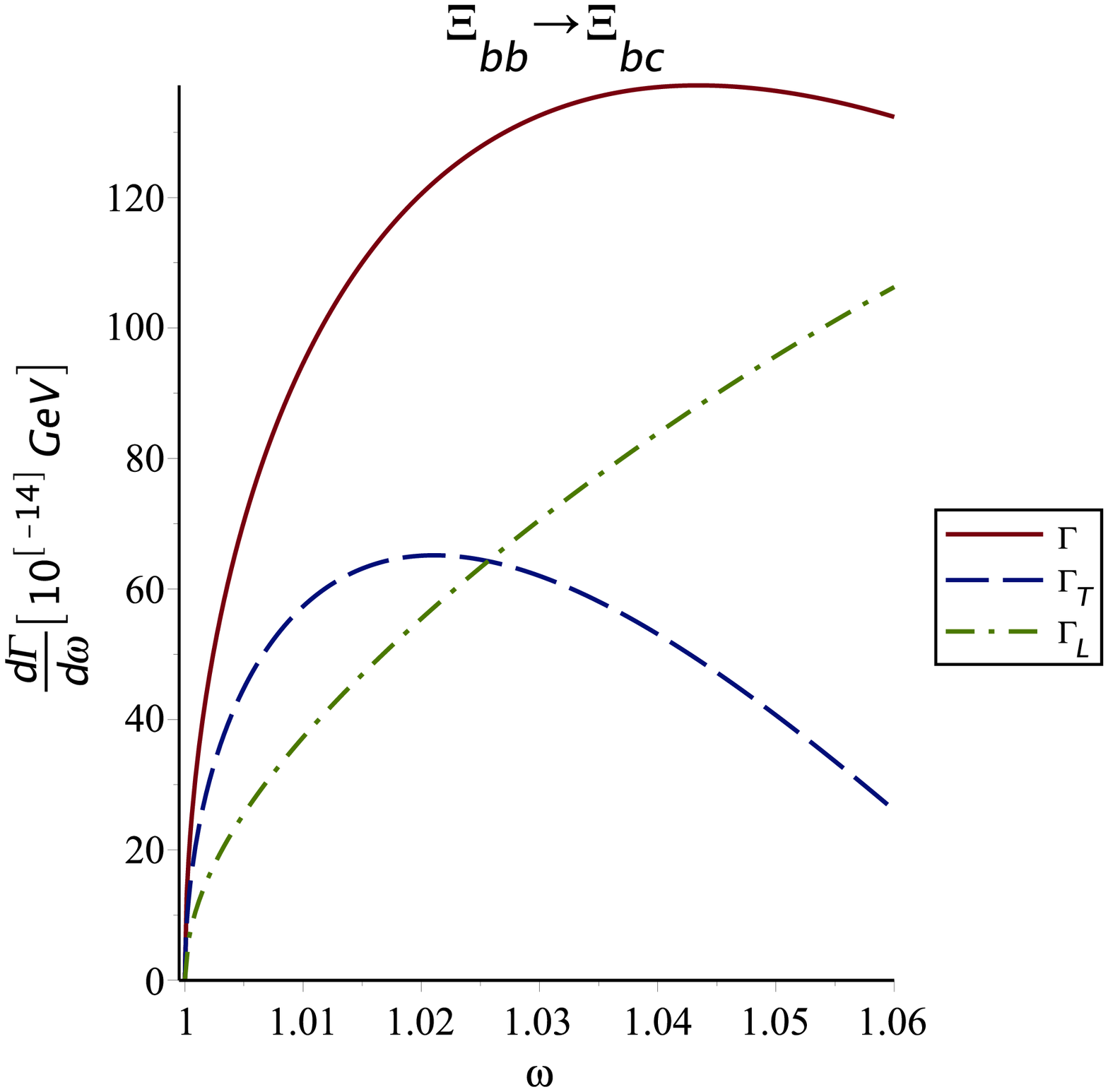}

\caption{$ \frac{d\Gamma_T}{d\omega} $, $  \frac{d\Gamma_L}{d\omega} $ and $ \frac{d\Gamma}{d\omega} $ semileptonic decay widths  in units of $ 10^{-14}$ GeV for  $\Xi_{bb}\rightarrow \Xi_{bc}\ell \bar {\nu}_{\ell}$ transition with $ \Lambda_B=2.5 $ GeV and $\ell=e$ or $\mu$. The results for $ \Omega_{bb} $ decay (not shown) are very similar. }
\label{fig:1}
\end{figure}

\begin{figure}[htbp]
\centering
\includegraphics[width=0.4\textwidth]{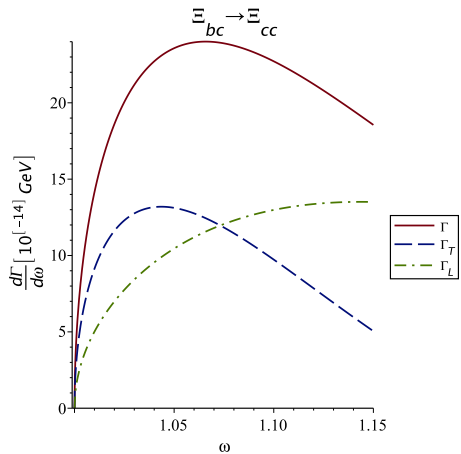}

\caption{$ \frac{d\Gamma_T}{d\omega} $, $  \frac{d\Gamma_L}{d\omega} $ and $ \frac{d\Gamma}{d\omega} $ semileptonic decay widths  in units of $ 10^{-14}$ GeV for  $\Xi_{bc}\rightarrow \Xi_{cc}\ell \bar {\nu}_{\ell}$ transition with $ \Lambda_B=2.5 $ GeV  and $\ell=e$ or $\mu$. The results for $ \Omega_{bc} $ decay (not shown) are very similar. }
\label{fig:2}
\end{figure}

The $ Q $ values of the semileptonic  decays are not large enough to allow for the semileptonic $ \tau $ modes. On the other hand, the $ Q $ values are sufficiently large to allow one to neglect the lepton masses in the semileptonic $ e $ and $ \mu $ modes.\\

The behavior of differential decay widths $ \frac{d\Gamma_T}{d\omega} $ and $  \frac{d\Gamma_L}{d\omega} $ with dependence on  $ \omega $ are shown in figures  \ref{fig:1} and \ref{fig:2}  for the  $\Xi_{bb}\rightarrow \Xi_{bc}$ and $\Xi_{bc}\rightarrow \Xi_{cc}$ transitions. We find that the results for doubly heavy $ \Xi $ and $ \Omega $ decays (not shown) are very close to each other.\\

In table \ref{tab:semileptonicLT} our results for the semileptonic decay widths (transverse $ \Gamma_T$,  longitudinal $ \Gamma_L $ and total $ \Gamma $) are presented. In table \ref{tab:semileptonic} we compare our results to the ones calculated in different models \cite{Albertus2008,Faessler2009,Ebert2004,hassanabadi2020,Ghalenovi:PLB2022}. \\
For semileptonic decays of doubly heavy baryons no experimental data are available at present.
From the numerical results in table \ref{tab:semileptonic} we can see that the decay widths are around the order $ 10^{-13}\sim 10^{-14}s^{-1} $. Different forms for the Isgur-Wise function may result in different decay widths. However, the order is not changed. 
Also the results are insensitive to the size parameter $ \Lambda_B $. The authors of Refs. \cite{Ghalenovi:PLB2022} and \cite{Faessler2009} have allowed the parameter $ \Lambda_B $ to vary in the range $2.5\leq \Lambda_B\leq 3.5$ GeV. Then, there are some uncertainties in their works. The error bars in the third and fourth columns of table \ref{tab:semileptonic} shows the variation of the size parameter $ \Lambda_B $ in their calculations which are bigger than the variations repored in Ref. \cite{Albertus2008}. Note that a smaller value of $ \Lambda_B $  gives smaller decay widths and \textit{vice versa}. In our calculations we take $\Lambda_B=2.5$ GeV \cite{hassanabadi2020,Ivanov:PRD1997,Faessler:PRD2006}. 
Regarding  $\Lambda_B=3.5$ GeV our values of decay widths would be larger by factor of $ \sim1.8 $ and $ \sim1.4 $ for  $bb\rightarrow bc$ and  $bc\rightarrow cc$ transitions respectively. \\
The authors of Refs. \cite{Albertus2005} and  \cite{Albertus2008} have studied the semileptonic decays of doubly heavy baryons in a non-relativistic quark model and evaluated the hadronic matrix elements parametrized in terms of six form factors. In the present work we follow their strategy. We develop their approach in the heavy quark limit and provide a way to simplify their evaluation of the hadronic matrix elements and form factors. From the last column of table \ref{tab:semileptonic} one can see that our results agree with their predicted decay widths. Our results of predictions for $bb\rightarrow bc$ transitions are lower than the estimates in Ref. \cite{Albertus2008}, while for $bc \rightarrow cc$ transitions our $ \Gamma $ values are higher than theirs. The same is for the estimates presented in the relativistic quark model \cite{Ebert2004}. On the other hand, the predictions of Ref. \cite{Faessler2009} in a relativistic covariant quark model are lower than ours. In comparison to Ref. \cite{hassanabadi2020}, our results are in good agreement with the ones presented in.\\

\begin{table} [h]
    \caption{Semileptonic decay widths of doubly heavy baryons in units of $10^{-14}$ GeV. $ \Gamma_T$ and $ \Gamma_L $ stand for the transverse and longitudinal contributions to the width $ \Gamma $. $ l $  stands for a light lepton, $ l=e, \mu $.}
      \label{tab:semileptonicLT}
    \centering
    \begin{center}
    \begin{tabular}{cccc|cccc}\hline

     ~~Decay~~& ~~$ \Gamma_T $~~&~~  $ \Gamma_L $~~ &~~ $ \Gamma $~~& ~~Decay~~& ~~$ \Gamma_T $~~&  ~~$ \Gamma_L $~~ & ~~$ \Gamma $~~\\
\hline
$\Xi_{bb}\rightarrow \Xi_{bc}\ell \bar {\nu}_{\ell}$ & 0.55 & 0.42 & 0.98&$\Omega_{bb}\rightarrow \Omega_{bc}\ell \bar {\nu}_{\ell}$ &0.58 & 0.45  &1.03\\
$\Xi_{bc}\rightarrow \Xi_{cc}\ell \bar {\nu}_{\ell}$& 1.32& 1.75  &3.08&$\Omega_{bc}\rightarrow \Omega_{cc}\ell \bar {\nu}_{\ell}$&1.40 &1.91   &3.32\\
\hline
    \end{tabular}
    \end{center}
\end{table}

\begin{table} [H]
    \caption{ Semileptonic decay widths of doubly heavy baryons in units of $10^{-14}$ GeV. $ l $  stands for a light lepton, $ l=e, \mu $.}
      \label{tab:semileptonic}
    \centering
    \begin{center}
    \begin{tabular}{ccccccc}\hline

     Decay& Our results&NRQM\cite{Ghalenovi:PLB2022}&RCQM\cite{Faessler2009}&NRQM\cite{hassanabadi2020}&RQM\cite{Ebert2004}&NRQM\cite{Albertus2008} \\
\hline
$\Xi_{bb}\rightarrow \Xi_{bc}\ell \bar {\nu}_{\ell}$ &0.98 & $1.75\pm 0.73 $& $0.80\pm 0.30 $&  0.94  & 1.63& 1.92$ ^{+0.25}_{-0.05} $\\
$\Xi_{bc}\rightarrow \Xi_{cc}\ell \bar {\nu}_{\ell}$&  3.08 & $ 4.39\pm0.83 $&$2.10\pm 0.70 $& 3.01 & 2.30 &2.57$ ^{+0.26}_{-0.03} $\\
$\Omega_{bb}\rightarrow \Omega_{bc}\ell \bar {\nu}_{\ell}$ &1.03  & $ 1.87\pm 0.76 $ &$0.86\pm 0.32 $ &0.99   & 1.70&2.14$ ^{+0.20}_{-0.02} $ \\
$\Omega_{bc}\rightarrow \Omega_{cc}\ell \bar {\nu}_{\ell}$&3.32 & $4.70\pm0.83$&$1.88\pm 0.62 $ & 3.28  & 2.48&2.59$ ^{+0.20} $ \\
\hline
    \end{tabular}
    \end{center}
\end{table}

By using the relation  $ Br=\Gamma\times\tau $ one can get the branching ratios of double heavy baryons where, $ \tau $ is the lifetime of the initial baryon. We use   $ \tau_{\Xi_{bb}}=370\times 10^{-15}s$, $  \tau_{\Xi_{bc}}=244 \times 10^{-15}s $  \cite{karliner2014}, $  \tau_{\Omega_{bc}}=220 \times 10^{-15}s $ and $  \tau_{\Omega_{bb}}=800 \times 10^{-15}s $ \cite{Kiselev2002,Kiselev2002b}.\\
 Our evaluated results are presented in table \ref{Beanching} and compared to those of other works \cite{Wang2017,hassanabadi2020,Ghalenovi:PLB2022}.
Our calculations for the branching ratios of $ bc \rightarrow cc $  decays are in good agreement with the ones reported by Ref. \cite{hassanabadi2020}. The predictions presented in Refs. \cite{Wang2017} and \cite{Ghalenovi:PLB2022} are higher than ours for all the transitions especially, for  $\Xi_{bb}\rightarrow \Xi_{bc}\ell \bar {\nu}_{\ell}$ and $\Omega_{bb}\rightarrow \Omega_{bc}\ell \bar {\nu}_{\ell}$ decays the predicted ratios given in Ref. \cite{Wang2017} exceed our estimates by a factor of  $\sim3.5$. 
\begin{table}[H]
    \caption{ Branching fractions of semileptonic decay widths for double heavy baryons.}
      \label{Beanching}
    \centering
    \begin{center}
    \begin{tabular}{ccccc}
  
    \hline
     Process  & Our results &\cite{hassanabadi2020} & \cite{Wang2017}& \cite{Ghalenovi:PLB2022} \\
\hline
$\Xi_{bc}\rightarrow \Xi_{cc}\ell \bar {\nu}_{\ell}$&1.14 $\times10^{-2}$ &$1.11\times10^{-2}$&$1.67\times10^{-2}$& $1.63\times10^{-2}$\\
$\Xi_{bb}\rightarrow \Xi_{bc}\ell \bar {\nu}_{\ell}$& 0.55$\times10^{-2}$ &$0.28\times10^{-2}$&$1.86\times10^{-2}$& $0.98\times10^{-2}$\\
$\Omega_{bc}\rightarrow \Omega_{cc}\ell \bar {\nu}_{\ell}$&1.10$\times10^{-2}$ &$1.10\times10^{-2}$&$1.32\times10^{-2}$&$1.57\times10^{-2}$\\
$\Omega_{bb}\rightarrow \Omega_{bc}\ell \bar {\nu}_{\ell}$& 1.26$\times10^{-2}$&&$4.49\times10^{-2}$& $2.27\times10^{-2}$\\
\hline
    \end{tabular}
    \end{center}
\end{table}
One notes that the rates for the exclusive semileptonic decay modes  $bb\rightarrow bc$ and  $bc\rightarrow cc$  are  rather small when compared to the total semileptonic inclusive rate. The remaining part of the inclusive rate would be filled in by decays into excited or multi-baryonic states.\\
In table \ref{tab:asymmetry} we compile our calculations for the averaged angular asymmetries  $ \alpha^{\prime} $ and $ \alpha^{\prime\prime} $, as well as the $ R_{L/T}=\Gamma_L/\Gamma_T $ ratio.
From table \ref{tab:asymmetry} one can see that the values of  averaged angular asymmetries for the $\Xi_{bb}$ ($ \Xi_{bc} $) and $\Omega_{bb}$ ($\Omega_{bc} $) semileptonic transitions are almost the same. That means the mean values of asymmetry parameters are independent of the light quark flavors ($u, d$ and $s$) inside the considered baryons. 
On the other hand, by increasing the size parameter $ \Lambda_B $, the mean values of asymmetry parameters increases. The increasing of  $ R_{L/T}$ for  $ bb \rightarrow bc  $ and $ bc \rightarrow cc $ transitions would be about  $ 20 \% $ and $ 40 \% $ respectively in the region $ \Lambda_B=2.5\sim 3.5$ GeV.
\begin{table} [H]
    \caption{Averaged values of the asymmetry parameters $ \alpha^{\prime} $ and $ \alpha^{\prime\prime} $ and  $ R_{L/T}$ ratio.}
      \label{tab:asymmetry}
    \centering
    \begin{center}
    \begin{tabular}{cccc|cccc}\hline

     ~~Decay~~& ~~$ \left\langle \alpha^{\prime}\right\rangle  $~~&~~  $ \left\langle \alpha^{\prime\prime}\right\rangle $~~ &~~ $ R_{L/T} $~~& ~~Decay~~& ~~$  \left\langle \alpha^{\prime}\right\rangle$~~&  ~~$ \left\langle \alpha^{\prime \prime}\right\rangle $~~ & ~~$ R_{L/T} $~~\\
\hline
$\Xi_{bb}\rightarrow \Xi_{bc}\ell \bar {\nu}_{\ell}$ & -0.05 & -0.21 &0.76 &$\Omega_{bb}\rightarrow \Omega_{bc}\ell \bar {\nu}_{\ell}$ &-0.05 &  -0.21 & 0.77\\
$\Xi_{bc}\rightarrow \Xi_{cc}\ell \bar {\nu}_{\ell}$&-0.09 & -0.45  &1.32&$\Omega_{bc}\rightarrow \Omega_{cc}\ell \bar {\nu}_{\ell}$&-0.08 & -0.46  & 1.35\\
\hline
    \end{tabular}
    \end{center}
\end{table}
\section{Conclusions} \label{Conclusions}
In summary, we have evaluated the mass spectra, magnetic moments and semileptonic decays of the ground-state doubly heavy $\Xi$ and $\Omega$ baryons. The calculations have been done in the framework of a non-relativistic quark model with the use of hypercentral approach. Introducing a simple form of the universal function we could simplify the weak transition form factors in the heavy quark limit. Finally, we have investigated the semileptonic decay rates and branching ratios driven by a $ b\rightarrow c $ quark transition for the $ J=\frac{1}{2} $ baryons. We have worked near to zero recoil point and in the heavy quark limit, with lepton mass neglected.  The results for doubly heavy $ \Omega $ decays are almost identical to the corresponding ones for doubly heavy $ \Xi $ decays. Since we have two heavy quarks, the light quark acts as a spectator and makes the results almost independent of the light quark mass. According to Ref. \cite{Faessler2009} the value of the size parameter $ \Lambda_B $ vary in the range $ 2.5\leqslant \Lambda_B\leqslant 3.5$ GeV. We have chose $ \Lambda_B=2.5 $ GeV. The decay widths obtained in Ref. \cite{Ghalenovi:PLB2022} are larger than ours for all transitions. If we choose $ \Lambda_B=3 $ GeV, the evaluated decay rates and branching ratios will be very close to the calculations of Ref. \cite{Ghalenovi:PLB2022}.
 A comparison between our results and those of other theoretical studies shows that our results are acceptable. Note that the triplet $ (J=\frac{3}{2}) $ doubly heavy baryons are dominated by the strong or electromagnetic decays. If these excited states be the initial ones, due to the smallness of the weak coupling, the weak decays can not be observed in the experiments. Therefore, one can neglect the calculations for the semileptonic decays of $ J=\frac{3}{2} $ doubly heavy baryons, since it is hard to be done in the experiments. We hope our results can be used to extract the CKM parameters $ V_{cb} $ from future experiments on the semileptonic decays of doubly heavy baryons. \\
\section{Acknowledgements}
This work was supported by Kosar University of Bojnord with the Grant number (No. 0011191444).

\end{document}